\documentclass{aa}
\usepackage{graphicx}
\usepackage{txfonts}
\usepackage{natbib}
\usepackage{aas_macros}
\bibpunct{(}{)}{;}{a}{}{,}
\begin{document}
   \title{On the cyclotron line in \hbox{Cep X-4}}

   \author{V.A.~McBride\inst{1}
          \and J.~Wilms\inst{2,3}
	  \and I.~Kreykenbohm\inst{4}
	  \and M.J.~Coe \inst{1}
	  \and R.E.~Rothschild \inst{5}
	  \and P.~Kretschmar \inst{6}
	  \and K.~Pottschmidt \inst{5}
	  \and J.~Fisher\inst{3}
	  \and T.~Hamson\inst{3}
          }

   \offprints{V.A.McBride \\\email{vanessa@astro.soton.ac.uk}}

   \institute{School of Physics \& Astronomy, University of Southampton,
     Highfield, SO17 1BJ, UK
              \and Dr.\ Karl Remeis-Sternwarte, Astronomisches
              Institut der Universit\"at Erlangen-N\"urnberg,
              Sternwartstr. 7, 96049 Bamberg, Germany 
               \and Department of Physics, University of Warwick,
               Coventry, CV4 7AL, UK 
	       \and INTEGRAL Science Data Centre, 16 ch.\ d'\'Ecogia,
               1290 Versoix, Switzerland
	       \and Center for Astrophysics and Space Sciences,
               University of California at San Diego, La Jolla, CA
               92093-0424, USA
              \and European Space Astronomy Centre (ESAC),
             European Space Agency, P.O.\ Box 50727, 28080, Madrid,
    Spain
             }

   \date{Received date / Accepted date}

   \abstract{Accreting X-ray pulsars provide us with laboratories for
     the study of extreme gravitational and magnetic fields, hence accurate descriptions of their observational properties contribute to our understanding of this group of objects.}{We aim to
     detect a cyclotron resonance scattering feature in the Be/X-ray
     binary \object{\hbox{Cep~X-4}} and to investigate pulse profile and spectral changes through the outburst.}{Spectral fitting and timing analysis are employed to probe the properties of \object{\hbox{Cep~X-4}} during an outburst in 2002 June.}{A
     previously announced cyclotron feature at 30.7\,keV is confirmed, while the source shows spectral behaviour and luminosity related changes similar to those observed in previous outbursts.  The
     long-term X-ray lightcurve shows a periodicity at 20.85\,d, which
     could be attributed to the orbit in this Be system.}{} 

   \maketitle
%

\section{Introduction}
Cep~X-4 was discovered as a transient source in 1972 June/July with
\emph{OSO 7} \citep{UlmerBaityWheaton1973}.  In 1988 March it was
detected with \emph{Ginga} during a month long X-ray outburst
\citep{Makino1988} and labelled \object{GS 2138+56}.  During this
outburst, where the source reached a maximum intensity of 100\,mCrab,
pulsations at $66.2490\pm0.0001$\,s were detected
\citep{KoyamaKawadaTawara1991,Makino1988b} as well as a cyclotron
resonance feature at $30.5\pm0.4$\,keV
\citep{MiharaMakishimaKamijo1991}, leading to its classification as an
X-ray binary pulsar.  Using the \emph{ROSAT} Position Sensitive
Proportional Counter, \citet{SchulzKahabkaZinnecker1995} detected
\hbox{Cep~X-4} in quiescence in 1993 January and in outburst in 1993 June.
During outburst, the pulse period was measured at
$66.2552\pm0.0007$\,s.  This observation lead to a refinement of the
source position, and a Be star at
$\alpha_{2000}=21^\mathrm{h}39^\mathrm{m}30\fs{}6$ and
$\delta_{2000}=+56\degr59'12\farcs{}9$ was proposed as the optical
counterpart \citep{RocheGreenHoenig1997}.  Long slit optical
spectroscopy of this proposed counterpart
\citep{Bonnet-BidaudMouchet1998} confirmed \hbox{Cep~X-4} as a Be/X-ray
binary with a distance of $3.8\pm 0.6$\,kpc.  In 1997 July and August, another outburst took place, this
time observed by the Burst and Transient Source Experiment
(\emph{BATSE}) and the Rossi X-ray Timing Explorer (\emph{RXTE}).
\citet{WilsonFingerScott1999} measured the pulsar spin down rate and
placed constraints on the orbital period using \emph{BATSE} and
\emph{RXTE} data from the 1993 and 1997 outbursts.  The most recent
outburst, occurring in 2002 June and lasting approximately one month,
was observed by \emph{RXTE} only.  

In this paper, we describe the spectral and timing analysis of pointed
\emph{RXTE} observations during the 2002 June outburst. In Sect.~2 we
present the observations and describe the reduction techniques.
Section~3 presents a possible orbital period in the long term lightcurve
while in Sect.~4 we fit X-ray spectra from the 2002 outburst.  In Sect.~5 the
pulse profiles are introduced.  Section~6 presents a discussion and
interpretation of the data, while the results are summarised in Sect.~7.

\section{Observations and Data Reduction}

\begin{figure}
\includegraphics[width=9cm]{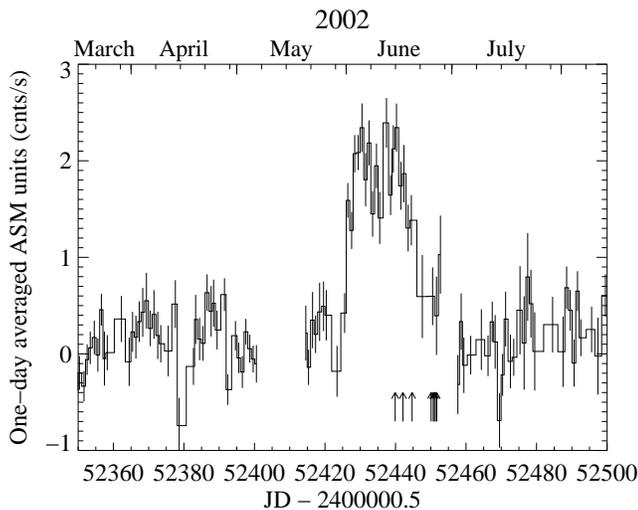}
\caption{The ASM lightcurve of \hbox{Cep~X-4} during the 2002
  June outburst, re-binned to a resolution of 1\,d and taking only
  one-day averages comprised of 10 or more dwells into account. Arrows
  indicate the dates of pointed \emph{RXTE} observations.  }
              \label{FigAsm}
\end{figure}

Figure~\ref{FigAsm} shows the \emph{RXTE} All Sky Monitor (\emph{ASM})
X-ray lightcurve of the month long outburst of \hbox{Cep~X-4} during the second quarter of 2002.  Observations with \emph{RXTE}'s pointed instruments were
performed during the outburst (arrows in Fig.~\ref{FigAsm} and
Table~\ref{TabObs}).

\begin{table}
\caption{\emph{RXTE} Observations of \hbox{Cep~X-4} during 2002 June
  outburst.  Luminosities are in the 2--10\,keV range and are calculated using a distance of $3.8\pm0.6$\,kpc \citep{Bonnet-BidaudMouchet1998}.}
\label{TabObs}
\begin{tabular}{lcrcl}
\hline
\hline
\noalign{\smallskip}
Obs ID & MJD & Exposure  & Luminosity & Period\\
70068-11- & & s & $10^{36}\mathrm{erg\,s}^{-1}$ & s \\
\hline
\noalign{\smallskip}
01-00 & 52439.9 & 800     & 1.40 & 66.41(0.05)\\
01-01 & 52442.1 & 816     & 1.06 & 66.36(0.04)\\
01-02 & 52444.7 & 880     & 0.87 & 66.25(0.04)\\
02-01 & 52450.1 & 1552    & 0.41& 66.35(0.02)\\
02-000 & 52450.7 & 19600  & 0.35 & 66.30(0.01)\\
02-00 & 52451.0 & 1456    & 0.31 & 66.33(0.03)\\
02-02 & 52451.3 & 8432    & 0.32 & 66.30(0.01)\\
02-03 & 52451.7 & 7088    & 0.27 & 66.296(0.004)\\

\hline
\end{tabular}

\end{table}

To increase the signal-to-noise we extracted data from only the top
layer of the  Proportional Counter Array
\citep[PCA][]{JahodaMarkwardtRadeva2006}.  As most of the source
photons, but only half the instrumental background, is detected in
this layer it is a good choice for a relatively weak source such as
\hbox{Cep~X-4}.  These data are \emph{Standard 2} mode data with 16\,s time resolution and 128 channel energy
resolution, and were employed to generate phase averaged spectra in
the range 3.5--20\,keV.  We added systematic errors of 0.1$\%$ of the
count rate in quadrature to all spectral bins across the energy range,
as indicated by near contemporaneous fits to the Crab spectrum.  For
pulse profiles and timing analysis, top layer \emph{Good Xenon} data
with a time resolution of 0.03\,s were used.

For all spectra we used the E\_$8\mu s$\_256\_DX1F mode High Energy
X-ray Timing Experiment \citep[HEXTE][]{RothschildBlancoGruber1998}
data, which has a temporal resolution of 8\,$\mu$s.  In all cases,
the signal-to-noise ratio was increased by adding together data from
both HEXTE clusters.  Data above 70\,keV were not used due to rapid
deterioration of the signal-to-noise at higher energies.

Analysis was done with \emph{HEADAS} version 6.0.4 and spectral
fitting with \emph{XSPEC} version 11.3.1w \citep{Arnaud1996}.


\section{Long-term lightcurve}
\begin{figure}
\centering
\resizebox{\hsize}{!}{\includegraphics{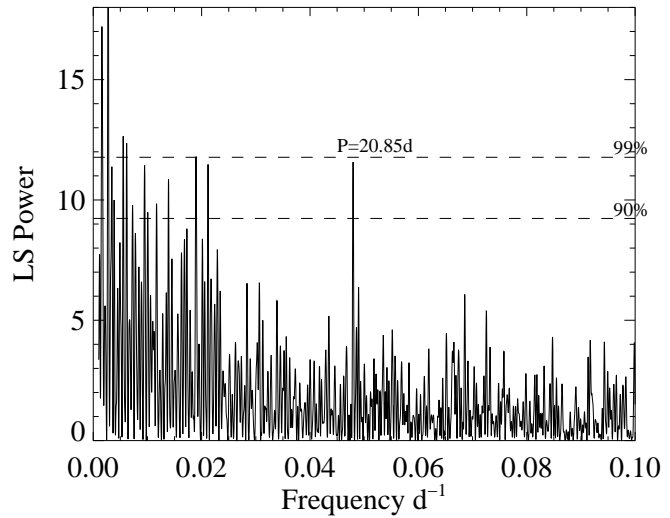}}
\caption{Lomb-Scargle periodogram of the ASM lightcurve of \hbox{Cep~X-4}}
\label{FigPdg}
\end{figure}

The archival \emph{RXTE}-ASM data from MJD 50090 to MJD 53873 were filtered on
the criterion that one-day averages should consist of 20 or more
dwells to make a long-term lightcurve of time resolution one day.
Without losing too many data points, this filtering gets rid of the
very noisiest points, but includes data from both the 1997 and 2002 outbursts of \hbox{Cep~X-4}.   Figure~\ref{FigPdg} shows the Lomb-Scargle
\citep{Lomb1976,Scargle1982} periodogram of this X-ray lightcurve.  To
determine the significance levels of peaks in the periodogram, 10\,000
Monte Carlo white noise simulations were generated
with the same mean, variance and sampling as that of the ASM data.
The 90\% and 99\% significance levels are shown in Fig.~\ref{FigPdg}.
A peak corresponding to a period at 20.85\,d, with an error of 0.05\,d is found above the 90\%
significance level. 

Apart from this periodicity and red noise
at low frequencies, no other significant long-term periodicities are
found in 10--1000\,d interval.  This periodicity is not related to any
systematic periods often present in ASM data (see
\citealt{FarrellONeillSood2005,Benlloch2004}). We speculate that this
period is caused by the orbit of the neutron star around its Be
companion.  It is very close to the lower limit of 23\,d as proposed by 
\citet{KoyamaKawadaTawara1991}, assuming a circular orbit for the binary system.  If the 1997 and 2002 outbursts are removed from the lightcurve, the 20.85\,d period is still present, albeit with a reduced significance.  The fact that inclusion of the outburst data boosts the signal is expected from a Be/X-ray binary, where it is common for outbursts to occur near periastron passage of the neutron star.  However, if the orbital period of this system is 20.85\,d then it is clear that we do not see bright outbursts at every periastron passage.  After its
initial discovery in 1972, \hbox{Cep~X-4} went into outburst 
every four to five years (1988, 1993, 1997, 2002).  If this is a
continuing trend,  we may expect to see an outburst in 2007.

\section{Spectral Analysis}

All observations from Table~\ref{TabObs} were added to create the
average outburst spectrum.

In the 3.5--70\,keV energy range we used a power law with Fermi-Dirac
cutoff to describe the continuum.  This model has the analytical form
\begin{equation}
\mathrm{FDCO}(E)=AE^{-\Gamma}\frac{1}{1+e^{(E-E_\mathrm{cut})/E_\mathrm{fold}}}
\end{equation}
and differs from the high energy cutoff model in that the transition
from power law to exponential is smooth, i.e. without a discontinuity
of the derivative at the cutoff energy.  Although the high energy
cutoff and NPEX \citep{Mihara1995} models also produced adequate fits
to the continuum, the FDCO model provides both a lower $\chi^2_{\nu}$
over the energy range and smaller uncertainties on the model
parameters.

\begin{figure*}
\sidecaption
\includegraphics[width=12cm]{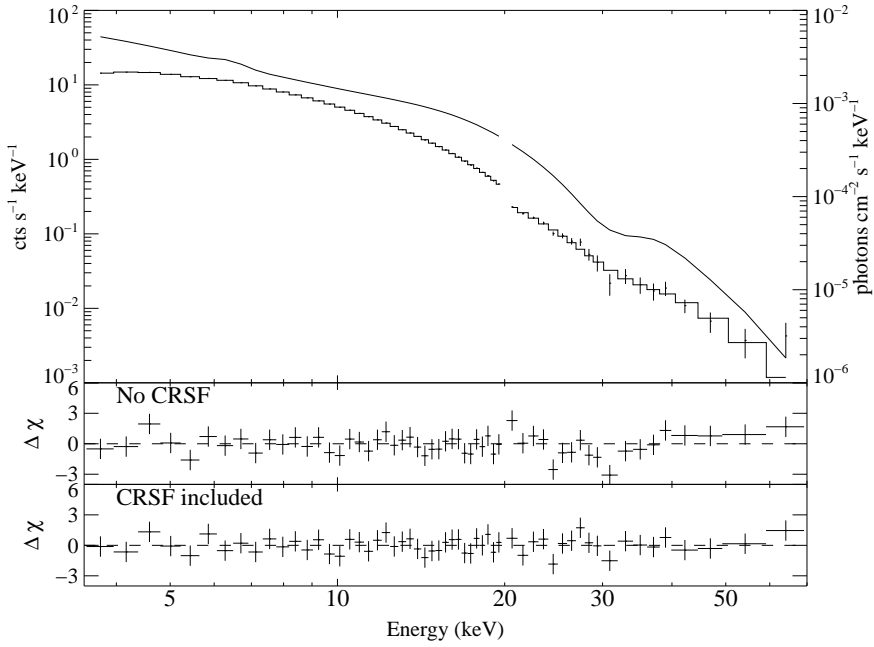}
\caption{The spectrum of \hbox{Cep~X-4} from 3.5--70\,keV.  The crosses show
  the data, the smooth curve shows the unfolded spectrum, and the
  histogram the model fit.  In the second panel the residuals, plotted
  as $\Delta\chi$, are shown for the case in which no CRSF is
  included in the model.  In the lower panel, these residuals are
  shown again -- this time including a CRSF in the model.}
              \label{FigPas}
\end{figure*}

The cutoff energy was fixed at 17\,keV from the fit with the lowest
$\chi^2$ as determined by
\citet{KoyamaKawadaTawara1991} during the 1988 outburst of \hbox{Cep~X-4}.
This approach was deemed reasonable, as preliminary fits to the data
converged at values close to this cutoff energy.  In addition we
modeled Fe K$\alpha$ emission with a Gaussian emission line, 
took into account photoelectric absorption using the \emph{phabs}
model in \emph{XSPEC}, with photoelectric cross-sections from
\citet{Balucinska-ChurchMcCammon1992} and element abundances from
\citet{AndersGrevesse1989}.  We added a $\sim$5\,keV wide emission line at $E_{\rm Gauss}\sim$14\,keV to
account for variation in the continuum over this energy range.  This
emission line is used to account for a feature noticed in spectra of a
number of accreting X-ray pulsars, e.g. \object{GS~1843+00} or
\object{Her~X-1} \citep{Coburn2001}.  It is not an instrumental
feature, as it has been observed with {\em RXTE}, {\em Ginga}
\citep{Mihara1995} and {\em BeppoSAX} \citep{sd98}, and should be
included in future models of pulsar X-ray continua.  An absorption
line with a Gaussian optical depth profile
\citep[Eq.~6]{CoburnHeindlRothschild2002} was used to model a weak
cyclotron resonant scattering feature (CRSF) at 30.7\,keV.  The
$F$-test probability of this improvement being by chance is
$1\times10^{-4}$.  See \citet{ProtassovVanDykConnors2002}, however,
for limitations of the $F$-test in these circumstances.  The average
outburst spectrum is shown in Fig.~\ref{FigPas} while the model
parameters are given in Table~\ref{TabSpecp}.

 \begin{table}
      \caption[]{Spectral parameters for the average outburst
      spectrum.  Errors indicate 90\% confidence intervals.}

         \label{TabSpecp}
	 \centering
         \begin{tabular}{ll}
            \hline\hline
            \noalign{\smallskip}
            \textrm{Parameter}      &  \textrm{Value}  \\
            \noalign{\smallskip}
            \hline
            \noalign{\smallskip}
	    $\Gamma$ & $1.44^{+0.08}_{-0.05}$\\
	    \noalign{\smallskip}
	    $E_\mathrm{cut}$ & 17 keV\\
	    \noalign{\smallskip}
	    $E_\mathrm{fold}$ & $10.9^{+0.8}_{-0.9}$ keV\\
	    \noalign{\smallskip}	
	    $N_{\mathrm H}$ & $2.1^{+0.2}_{-0.1}\times10^{22}\textrm{cm}^{-2}$\\
	    \noalign{\smallskip}
	    $E_\mathrm{c}$ & $30.7^{+1.8}_{-1.9}$ keV\\
	    \noalign{\smallskip}
	    $\sigma_\mathrm{c}$ & $3.6^{+2.9}_{-1.5}$ keV\\
	    \noalign{\smallskip}
	    $\tau_\mathrm{c}$ & $0.7^{+0.3}_{-0.2}$\\
	    \noalign{\smallskip}
	    $E_\mathrm{Fe}$ & $6.38^{+0.05}_{-0.06}$ keV\\
	    \noalign{\smallskip}
	    $\sigma_\mathrm{Fe}$ & $0.32^{+0.08}_{-0.10}$ keV\\
	    \noalign{\smallskip}
	    Fe EW & $73.3\pm0.1$ eV\\
	    \noalign{\smallskip}
	    $E_{\rm Gauss}$ & $14.4^{+0.2}_{-0.1}$ keV\\
	    \noalign{\smallskip}
	    Gauss EW & $4.6^{+5.2}_{-1.6}$ keV\\
            \noalign{\smallskip}
	    Flux (3.5--10\,keV) & $1.8^{+1.3}_{-0.9}$\,erg\,cm$^{-2}$\,s$^{-1}$\\
	    \noalign{\smallskip}
            $\chi^2_\nu$\,(dof) & 0.74\,(43)\\
	    \noalign{\smallskip}
            \hline
         \end{tabular}
 
\end{table}

\begin{figure}
\label{FigSpecchange}
\resizebox{\hsize}{!}{\includegraphics{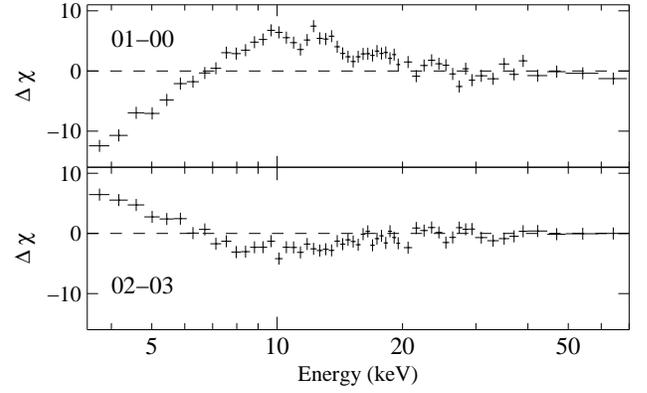}}
\caption{Residuals, plotted as $\Delta\chi$, as a function of energy
  for the average outburst model compared to the spectra of
  observations 01-00 and 02-03.}
\end{figure}

The average outburst spectrum is strongly influenced by the longest
(20\,ks) observation on MJD 52450.7.  In order to see how well this
spectrum represented individual observations we normalised the average
outburst model to each individual observation by multiplying it by a
constant.  This allows us to notice broad changes across the spectrum.
In earlier data (e.g., observations 01-00, 01-01) the model strongly
overestimates the data at the soft end ($<7$\,keV) of the spectrum.
Whereas for later observations (e.g. 02-02, 02-03) the soft end is
underestimated (see Fig.~\ref{FigSpecchange}).  To investigate this
effect we fitted the spectra of individual observations, this time
allowing the normalisation parameters, the power law index and
photoelectric absorption parameter to vary.  Our fits show a clear
steepening (from $1.19\pm 0.06$ to $1.50\pm0.04$) of the power law as
the outburst progresses, but the absorption column, although poorly
constrained by the model, is consistently high
($\sim2.5\pm0.3\times10^{22}$\,cm$^2$) through the outburst.
This suggests that the spectrum becomes softer as the outburst
progresses.  There is, however, little variation around the energy
range containing the cyclotron feature.

As cyclotron features are heavily dependent on the viewing angle and
geometry of the line-forming region, a phase resolved study of
\hbox{Cep~X-4} would be beneficial to understanding this system.
However, the low count statistics in the region around the cyclotron
line in this dataset make such a study unfeasible.
 
\section{Pulse Profiles}

\begin{figure}
\centering
\includegraphics[angle=0,width=9cm]{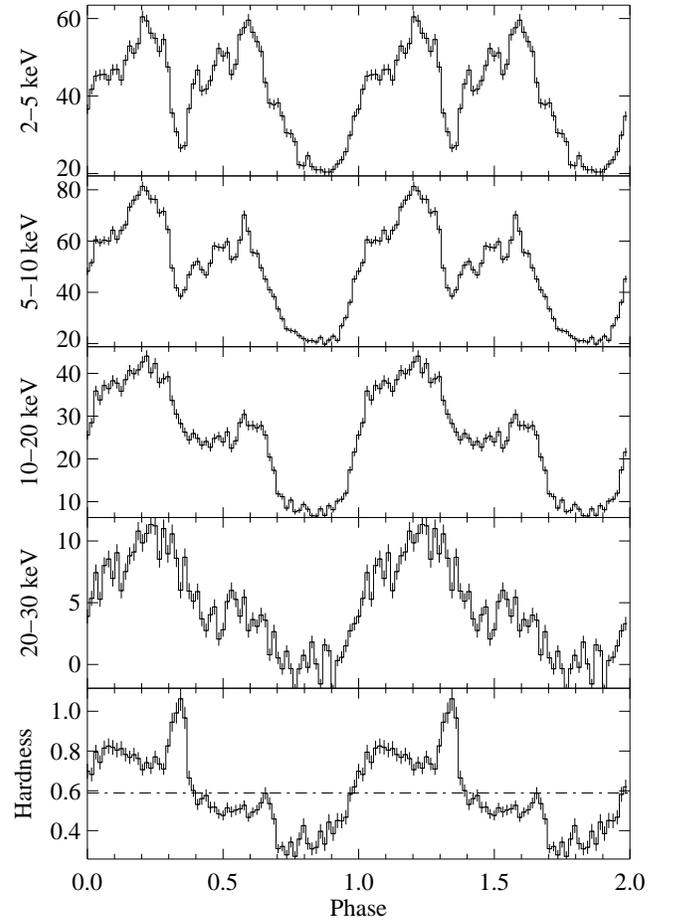}
\caption{Pulse profiles as a function of energy for observation 01-00
  (MJD 52439.9) shown in counts/s/PCU.  Profiles are plotted twice for
  clarity.  Hardness ratio is 10--20\,keV band divided by the
  2--5\,keV band.}
              \label{FigPulsep0100}
\end{figure}

\begin{figure}
\centering
\includegraphics[angle=0,width=9cm]{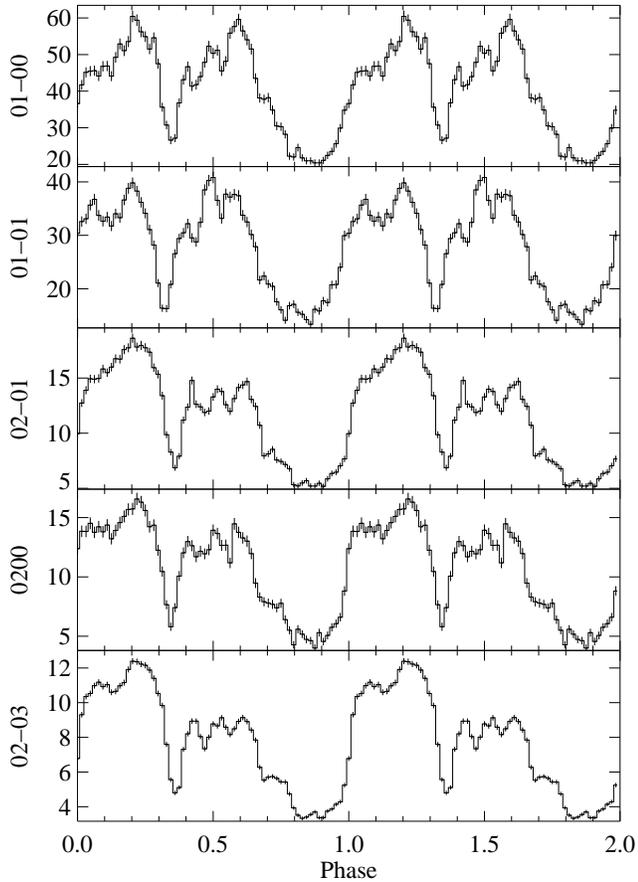}
\caption{Pulse profiles in the 2--5\,keV energy for a number of
  observations throughout the outburst.  Luminosity decreases from the
  top to bottom panel. (See Table~1 for luminosity values.)}
              \label{FigPulseAllSoft}
\end{figure}

An epoch folding search \citep{LeahyDarbroElsner1983} was used to
determine the source pulse period for each observation (see
Table~\ref{TabObs}).  The pulse profiles were generated by folding the
lightcurves at a period of 66.30\,s.  The pulse profiles are
double-peaked and complex through all energy bands
(Fig.~\ref{FigPulsep0100}).  Both of the peaks show additional
structure which is prominent at low energies and becomes less coherent
at higher energies.  The pulse profile in the 2--5\,keV energy changes
significantly as the source luminosity drops (see
Fig.~\ref{FigPulseAllSoft}) with the second peak becoming weaker
relative to the first as the outburst progresses.

\begin{figure}
\centering
\includegraphics[width=9cm]{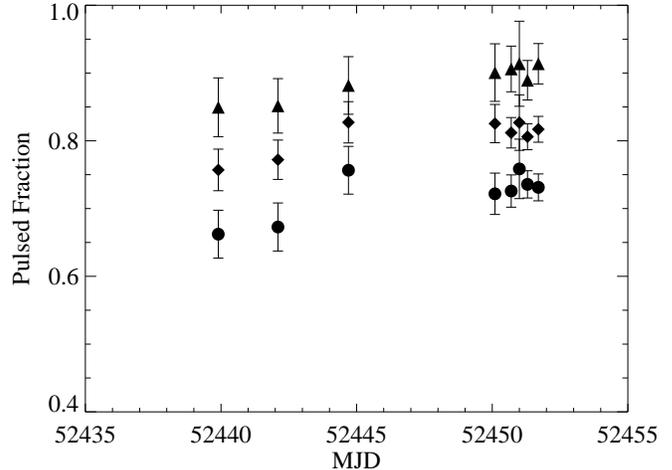}

\caption{Pulsed fraction through the outburst.
  Triangles represent the pulsed fraction in the 10--20\,keV band,
  diamonds for the 5--10\,keV band and circles for the 2--5\,keV
  band.}
\label{FigPulseFrac}
\end{figure}

The pulsed fraction, as defined by
$(F_\mathrm{max}-F_\mathrm{min})/F_\mathrm{max}$, is shown in
Fig.~\ref{FigPulseFrac} and indicates an overall increase in softer
bands (2--5\,keV and 5--10\,keV) as the outburst progresses.  This is
contrary to what was observed in the 1997 outburst, where
\citet{WilsonFingerScott1999} saw the pulsed fraction decreasing as
the source luminosity dropped.

\section{Discussion}

\subsection{Spectrum}

A cyclotron line is identified in the average outburst spectrum at an
energy of $30.7^{+1.8}_{-1.9}$\,keV, confirming the cyclotron line
discovered at this energy in the 1988 outburst of \hbox{Cep~X-4} by
\citet{MiharaMakishimaKamijo1991} with \emph{Ginga}.  In the 2002
outburst, as in 1988, the cyclotron feature appears as an absorption
line.  With the intention of comparing the line properties between the
2002 and 1998 outbursts we fit the 2002 data with the CYCLABS model,
which was used by \citet{MiharaMakishimaKamijo1991}.  The line
parameters for the 2002 outburst, using the CYCLABS model are
consistent, within errors, with those measured using the Gaussian
optical depth model and given in Table~\ref{TabSpecp}.  However, the
overall model fit is slightly worse.  We find values of the line
depth and width that are around a factor three smaller than those
measured in 1988 ($W_\mathrm{c}=15.0\pm1.4$\,keV and
$D_\mathrm{c}=2.93\pm0.10$, \citealt{MiharaMakishimaKamijo1991}), but
note that the cyclotron line was near the upper limit of the
\emph{Ginga} energy range. The observed energy of the cyclotron line is 
given by
\begin{equation}
\label{EqnCyclotron}
E_{\rm c}\simeq 11.6\,{\rm keV}\times \frac{1}{1+z}\times \frac{B}{10^{12}\rm{G}}
\end{equation}
Assuming that the observed feature is the fundamental ($n=1$) and
assuming a gravitational redshift $z = 0.3$ at the surface of a typical neutron star
mass of $1.4\,M_{\odot}$ and radius of 10\,km, we can obtain the
neutron star magnetic field to be
$B\simeq(3.4\pm0.2)\times10^{12}$\,G from Eq~\ref{EqnCyclotron}.

The Fe K$\alpha$ line has an equivalent width of 73\,eV which is
within the range of the Fe line equivalent width during the 1988
outburst, and is of the same order as that of Fe K$\alpha$ emission
lines observed in other accreting X-ray pulsars \citep{Nagase1989}.

The $N_{\mathrm H}$ column density is more than double that noted
before in any previous outburst of \hbox{Cep~X-4}
\citep{KoyamaKawadaTawara1991,SchulzKahabkaZinnecker1995} and also
exceeds that derived from the reddening of the optical counterpart
\citep{Bonnet-BidaudMouchet1998}.  We interpret this increase in the
column density as a local effect and ascribe it to possible increased mass
transfer through the accretion stream or warping of the accretion disc.

The spectrum certainly changes with source luminosity, becoming softer
as the luminosity fades.  A similar effect occurred in the 1988
outburst with \citet{KoyamaKawadaTawara1991} reporting a change in the
power law index from $1.10\pm0.01$ to $1.14\pm0.01$ over 10 days of
the fading outburst.  Although the change in power law slope occurs
over a similar timescale in this latest data set, the change is more
pronounced.  This is in contrast to \object{V0332+53}, an accreting
X-ray pulsar where the spectrum became harder through the decline of
the outburst \citep{MowlaviKreykenbohmShaw2006}.  More observations,
including data below 3\,keV, will be useful in assessing whether 
both the power law and \textsc{H~i} absorption change over the
outburst.

\subsection{Pulse Profiles}

For the 1997 outburst \citet{WilsonFingerScott1999} describe the pulse
profile using the model of \citet{BrainerdMeszaros1991} in which the
total pulse profile is made up of two components.  A hard
pencil beam is caused by photons from one magnetic pole being
upscattered through the accretion column.  These photons contribute to
the prominent peak in the hardness ratio around phase 0.3.  Simultaneously, a soft, double-lobed peak originates from the antipodal magnetic pole when
cyclotron photons are backscattered.  These photons are backscattered
into the neutron star or gravitationally focused around the neutron
star to form a fan beam of soft photons. 

The pulse profiles of \hbox{Cep~X-4} bear a striking resemblance to
the complex pulse profiles of \hbox{Vela~X-1}
\citep{KreykenbohmCoburnWilms2002} in the soft X-rays.  Both these
accreting systems show two main pulses each subsequently made up of
further structure.  As with \hbox{Vela~X-1} this complexity
disappears at higher energies leaving a clearly double-peaked profile.
Interpretation of the complexity is uncertain, ranging from variable
absorption over the neutron star spin phase \citep{NagaseHayakawaMakino1983} to anisotropic emission through the accretion column at the polar cap. 

The pulse profiles, especially those at lower energies, show clear
evolution over the outburst.  Over the 12 days covered by the {\em
  RXTE} pointed observations, the luminosity decreased by a factor of
5 and the two peaks in the double-pulsed profiles in the 2--5\,keV
band became distinctly more dissimilar to each other with the
flux in one peak decreasing far more sharply with
luminosity than the decrease in flux in the other pulse peak.  Comparable
changes were noted in the 1997 outburst of \hbox{Cep~X-4}
\citep{MukerjeeAgrawalPaul2000}, where the relative strengths of the
two pulses comprising the double-pulsed profile were reversed and the
interpulse became stronger with decreasing source luminosity.  

It is not uncommon for binary pulsars to show strong variations in their pulse profiles over the duration of an outburst.  For
\object{EXO~2030+375}, which also shows complex double-peaked
profiles, a reversal of the dominant pulse was detected
\citep{ParmarWhiteStella1989}.  A simple geometric model comprising
pencil and fan beams from two offset magnetic poles was employed.
This interprets the above reversal as a switching of the dominant
radiation from one magnetic pole to the other.
\citet{ParmarWhiteStella1989} also showed that as the luminosity
decreased by a factor of 100, the beam pattern changed from fan to
pencil beam.  \citet{BaskoSunyaev1976} propose that a fan beam will be the dominant
pulse shape at higher luminosities ($>10^{37}\,{\rm erg\,s}^{-1}$)
where the accreting matter forms a shock above the neutron star
surface and radiation escapes predominantly from the side of the
accretion column \citep{WangFrank1981}.  In lower luminosity
scenarios, the infalling matter may be decelerated by Coulomb
interactions at the neutron star surface
\citep{BaskoSunyaev1975,KirkGalloway1981}, giving rise to a pencil
beam emission pattern.

The pulse profiles in Fig.~\ref{FigPulsep0100}
and~Fig.~\ref{FigPulseAllSoft} show behaviour similar to V0332+53
\citep{TsygankovLutovinovChurazov2006} in its high luminosity
($\sim10^{38}$\,erg\,s$^{-1}$) states, i.e., a change in the relative
heights of the double peaked profile with increasing energy as well as
a marked change in the relative intensity of the peaks in the soft
bands (3--6\,keV and 6--10\,keV) as the source luminosity decreases.
Luminosities $> 10^{37}$\,erg\,s$^{-1}$ are certainly not seen in
this outburst of \hbox{Cep~X-4}, yet it shows similar changes of the pulse
profile.  This may indicate that the soft pulse profile evolution in this source could be attributed to changes in the flux contribution from the two magnetic poles, although the prominent feature at phase
0.3 in the hardness profile (Fig.~\ref{FigPulsep0100}) is not
obviously explained by such a scenario.  Accurate orbital parameters for \hbox{Cep~X-4}, which are not yet
determined, together with modeling of the pulse profiles will help to highlight
the geometry and clarify our understanding of the pulse profile behaviour.

\section{Summary}
\begin{enumerate}
\item We have confirmed the detection of a cyclotron line first noted
  by \citet{MiharaMakishimaKamijo1991} in the 1988 outburst of \hbox{Cep~X-4}.
\item We have observed a column density a factor of two higher than
  previously observed for this source.  We attribute this increased
  density to an effect local to the Be/X-ray binary system of \hbox{Cep~X-4},
  such as possible partial obscuration by the accretion stream onto the 
  neutron star.
\item We note changes in pulse profiles both with energy and with
  decreasing source luminosity.  Although similar changes are noticed
  in the pulse profiles of other accreting X-ray pulsars there exists
  no global interpretation of these effects.  Modeling of individual
  systems, incorporating the orbital parameters, can shed light on the
  pulsar geometry and emission patterns.
\item A softening of the source spectrum with decreasing luminosity,
  as was noted in a previous outburst of \hbox{Cep~X-4}
  \citep{KoyamaKawadaTawara1991} is observed, in contrast to the
  hardening noticed during the outburst decay of V0332+53.
\item A tentative orbital period of 20.85\,d for \hbox{Cep~X-4} is revealed in
  the long term X-ray lightcurve.
\end{enumerate}

\begin{acknowledgements}
VAM would like to acknowledge the NRF (S.Africa), the British Council
and Southampton University. 
\end{acknowledgements}

\bibliographystyle{aa}

\end{document}